\documentclass{aa}

\def\x2{$\chi^{2}$}

\begin{document}
   \title{On the use of photometric redshifts for X-ray selected AGNs}


  \titlerunning{On the use of photometric redshifts for X-ray selected AGNs}
   \authorrunning{S. Kitsionas et al.}

   \author{S. Kitsionas\inst{1,2}
          \and
          E. Hatziminaoglou\inst{3}
          \and 
          A. Georgakakis\inst{1}
          \and
          I. Georgantopoulos\inst{1}
          }

   \offprints{S. Kitsionas, \email{skitsionas@aip.de}}

   \institute{Institute of Astronomy \& Astrophysics,
              National Observatory of Athens, 
              I. Metaxa and V. Pavlou str., \\
              P. Penteli, 
              GR-15236, Athens, Greece
         \and
              Astrophysikalisches Institute Potsdam, 
              An der Sternwarte 16, D-14482, 
              Potsdam, Germany
         \and
              Instituto de Astrofisica de Canarias, C/ V\'{\i}a L\`{a}ctea,
              s/n, E-38205,  La Laguna, Tenerife, Spain
             }

   \date{Received ; accepted }

   \abstract{
In this paper we present photometric redshift estimates for a sample
of X-ray selected sources detected in the wide field ($\sim 2 \,
\rm deg^2$), bright  [$f_{X}(\rm 0.5-8\,keV)\approx10^{-14}\,
erg\,s^{-1}\,cm^{-2}$] XMM-{\it Newton}/2dF survey. Unlike deeper
X-ray samples comprising a large fraction of sources with colours
dominated by the host galaxy, our bright survey primarily probes the
QSO X-ray population. Therefore photometric redshift methods employing
both galaxy and QSO templates need to be used. We employ the
photometric redshift technique of Hatziminaoglou, Mathez \& Pell\'{o} 
(2000) using 5-band photometry from the SDSS. We separate our X-ray 
sources according to their optical profile to point-like 
and extended. We apply QSO and galaxy templates to the
point-like and extended sources respectively. X-ray sources 
associated with Galactic stars are identified and discarded from our 
sample on the basis of their unresolved optical light profile, their low
X-ray--to--optical flux ratio and their broad band colours that are best 
fit by stellar templates. Comparison of our
results with spectroscopic redshifts available, 
  allows calibration of our method and estimation of
the photometric redshift accuracy. For $\sim 70$ per cent of the
point-like sources photometric redshifts are correct within $\rm
\delta z \la 0.3$ (or $\sim 75$ 
per cent have $\rm \delta z/(1+z) \la 0.2$), 
and the rms scatter is estimated to be $\rm \sigma_z= 0.30$.
Also, in our {\it X-ray selected} 
point-like sample we find that about 7 per cent of the sources
have optical colours redder than those of {\it  optically} selected
QSOs. Photometric redshifts for these systems using existing QSO
templates are most likely problematic. 
 For the optically extended objects the photometric redshifts work only 
 in the case of red ($g - r > 0.5$ mag) sources yielding 
 $\rm \delta z \la 0.15$ and $\rm \delta z/(1+z) \la 0.2$
 for 73 and 93 per cent respectively.  
 The results above are consistent with 
 earlier findings on the application of combined galaxy/QSO 
 photometric redshift techniques in the {\it Chandra} Deep Field North.   
 However, we find that the above photometric redshift technique does not
 work in the case of extended sources with blue colours ($g-r<0.5$).
 Although these form a significant fraction of the extended sources
 ($\approx  40\%$), they cannot be fit successfully by QSO or galaxy templates, or any linear combination of the two.
   }

   \maketitle

\label{firstpage}

\section{Introduction}

In the past few years, X-ray surveys carried out by the XMM-{\it
Newton} and the {\it Chandra} observatories have significantly
improved our understanding of X-ray selected AGNs (Hasinger et
al. 2001; Brandt et al. 2001; Giacconi et al. 2001; Barger et
al. 2002; Brandt et al. 2002; Mainieri et al. 2002;  Georgantopoulos
et al. 2004). Moreover, serendipitous studies using the large volume
of archival data provided by these missions are well underway aiming
to further push our 
knowledge on both the X-ray properties and the evolution of AGNs
(Green et al. 2004; Kim et al. 2004a; 2004b).        

To maximise their scientific impact, the above large scale projects
require redshift information for a large number of X-ray
sources. For example the recently released XMM-{\it Newton}
Serendipitous Source Catalogue comprises about 50,000 sources with
more archival observations being accumulated. Spectroscopic follow-up
programs are underway (e.g. Barcons et al. 2002) but they are
expensive in telescope time and completion may take
years. Moreover, many sources are expected to be optically
faint rendering spectroscopic observations difficult even with 10-m
class telescopes.   

The issues above have motivated the development of photometric
redshift techniques that use multiwaveband imaging to estimate the
redshifts of extragalactic sources. These methods have become
increasingly popular over the last few years due to their
effectiveness and the relatively small investment in observing time
that they require compared to spectroscopy. For normal galaxies in
particular, with optical colours dominated by stellar processes,
photometric redshifts have proven exceptionally successful
achieving accuracies that allow detailed statistical
studies (e.g. Budavari et al. 2003; Csabai et al. 2003).

For AGN dominated systems with featureless continua, photometric
redshift estimates are more challenging. Nevertheless, significant
effort has been expended in this direction in recent years. Budavari
et al. (2001) and Richards et  al. (2001) used Sloan Digital Sky
Survey (SDSS) data to estimate photometric redshifts of {\it
optically} selected QSOs: their results are accurate within $\delta
\rm z<0.2$ for 70 per cent of their sample.  

For {\it X-ray} selected samples, Gonzalez \& Maccarone (2002) have 
estimated photometric redshifts for X-ray sources to the limit
$f_X(\rm 0.5-8\,keV)\approx10^{-16}\, erg \, s^{-1}\, cm^{-2}$. Using
galaxy templates these  authors find good agreement between
photometric and spectroscopic redshifts {\it only} for systems
dominated by light from the host galaxy rather than the central
engine. This class of sources is the dominant 
population representing about 90 per cent of their X-ray selected
sample to the flux limit above. Similarly, Mobasher et al. (2004) used 
optical data from
the GOODS survey to estimate photometric redshifts for X-ray selected
sources in the {\it Chandra} Deep Field North [$f_X(\rm 0.5 - 8 \, keV )
\approx 10^{-16}\, erg\, s^{-1} \, cm^{-2} $]. Although for most of
these X-ray faint sources galaxy templates provide adequate
photometric redshift estimates, they fail for powerful QSOs.  
Babbedge et al. (2004) recently presented a combined galaxy-QSO 
 template approach to derive photometric redshifts for the 
 X-ray sources in the {\it Chandra} Deep Field North (CDFN). 
 
 The use of combined galaxy-QSO templates is 
 imperative in the case of X-ray surveys at brighter limits which  comprise a
significant fraction of powerful AGNs with optical light dominated by
the central engine. In this paper, we address this issue providing  
 photometric redshifts for X-ray sources detected in a wide 
 field  ($\sim 2\,\rm deg^2$), shallow
[$f_X(\rm 0.5-8\,keV)\approx10^{-14}\, erg\, s^{-1} \, cm^{-2}$]
XMM-{\it Newton} survey near the North Galactic Pole region (part of
the XMM-{\it Newton}/2dF survey; Georgakakis et al. 2003, 2004). The
strength  of this sample is that it overlaps with the SDSS 
 (York et al.  2000) and hence, high quality 5-band
photometry is available  allowing photometric redshift estimates. Also
spectroscopic redshifts are available for a sub-sample of the X-ray 
source population from the 2dF QSO Redshift Survey (2QZ; Croom et
al. 2001), the SDSS survey (York et al. 2000; Stoughton  et
al. 2002) as well as our own spectroscopic campaign. The
spectroscopic redshifts allow both calibration of the photometric
redshift method and estimation of its accuracy. 

In the following section we give a detailed description of both the
X-ray and optical (photometric and spectroscopic) data used in the
present study. In section 3 we give a brief summary of the photometric
redshift method while section 4 presents our results. These are
discussed in section 5. Our conclusions are summarised in section 6. 

\section{The Sample}

The X-ray data are from the XMM-{\it Newton}/2dF survey, a wide area,
shallow [$f_{X}(\rm 0.5-8 \, keV) \approx 10^{-14}\,
erg\,s^{-1}\,cm^{-2}$] X-ray sample near the North and the  
South Galactic Pole regions. The X-ray data reduction, 
source extraction, flux estimation and catalogue generation 
are described in detail by Georgakakis et al. (2003, 2004).  

In the present study we concentrate on the XMM-{\it Newton}/2dF survey
sub-sample near the North Galactic Pole region. This is due to the
wealth of homogeneous follow-up optical photometric and spectroscopic 
observations available in this region. A total of 291 X-ray
sources have been detected in the 0.5-8 keV spectral band above the $5
\sigma$ threshold.  

The optical photometric data are from the SDSS Early Data Release 
(EDR; Stoughton et al. 2002). The SDSS is an ongoing imaging and
spectroscopic survey  that aims to cover about 10,000 deg$^{2}$ of the
sky. Photometry is performed in 5 bands ({\it ugriz}; Fukugita et
al. 1996) to  a limiting magnitude of $g \approx 23.0$ mag. These
data are used for the optical identification of the X-ray sample using
the  method described by Downes et al. (1986). We propose 193 optical
counterparts out of 291 X-ray sources.  

Spectroscopic data for the optically brighter X-ray sources are from
the SDSS and the 2QZ. The SDSS will obtain spectra for over 1
million objects, including galaxies brighter than $r=17.7$\,mag,
luminous red galaxies to z $=0.45$ and colour selected QSOs 
(York et al. 2000; Stoughton et al. 2002; Richards et al. 2002). 
The 2QZ is a large-scale spectroscopic campaign that fully exploits the
capabilities of the 2dF multifibre spectrograph on the 4m 
Anglo-Australian Telescope (AAT). This project provides high quality
spectra, redshifts and spectral classifications for 23,000 optically
selected $b_{j}<20.85$\,mag QSOs (Croom et al. 2001). In addition
to publicly available  spectroscopic data from the 2QZ and the
SDSS spectroscopic surveys, we use redshift measurements  from our
own on-going spectroscopic campaign of the XMM-{\it Newton}/2dF
sources. Part of this data are  presented by Georgakakis et
al. (2004).

\begin{figure}
\setlength{\unitlength}{1mm}
\begin{picture}(80,90)
\includegraphics{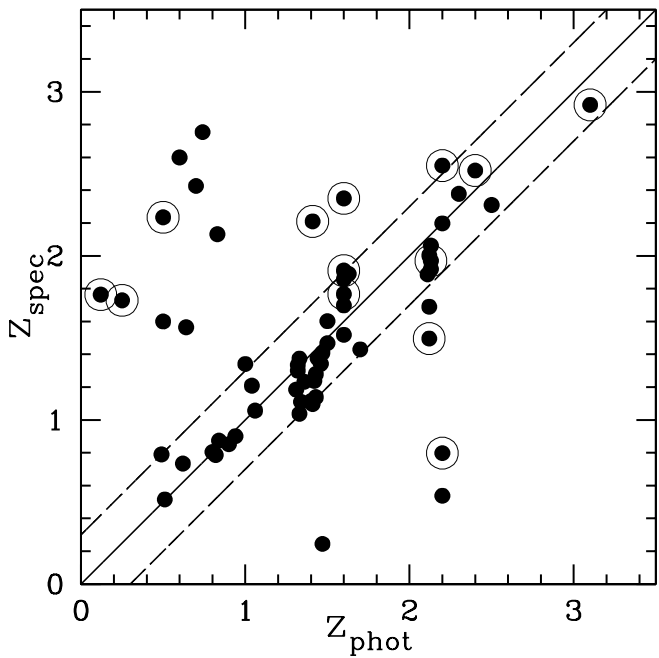}
\end{picture}

\caption{Comparison between z$_{spec}$ and z$_{phot}$ for X-ray  
sources with point-like optical light profile and spectroscopic
redshift measurements. A large circle on top of a symbol is for 
sources with $r > 21$. The solid line is the z$_{spec}$ = z$_{phot}$
locus. The dashed lines define the $\rm \delta z = \pm 0.3$ envelope around
the  z$_{spec}$ = z$_{phot}$ line.}  
\label{fig:one}
\end{figure}

\section{The photometric redshift method}

Photometric redshifts are estimated using the method described by 
Hatziminaoglou, Mathez \& Pell\'{o} (2000; HMP00). In brief this is
based on a standard $\chi^{2}$ minimisation technique where the
photometric redshift of an object is estimated by comparing its
multiband photometry with model QSO Spectral Energy Distribution (SED)
templates, shifted in redshift space and integrated through the
bandpass throughput functions. Redshift is allowed to vary in 
the range 0--6 with a linear step of 0.1. For each template and 
at each redshift, the $\chi^{2}$ probability of each source is 
calculated. Once a local minimum in $\chi^{2}$ is found, a 
refined search for a better solution is made around it, with a step 
of 0.01. More than one local minima might exist, as in all template 
fitting techniques. The photometric redshift of the source is then 
given as the redshift corresponding to the global minimum of 
these $\chi^{2}$ values.

The QSO templates are produced by varying the optical power-law
spectral index of simulated QSO spectra between 0 and 1, while keeping
the UV spectral index constant at 1.76 (Wang, Lu \& Zhou
1998). We note however, that varying the UV spectral index does not
modify our results and conclusions. Emission lines (Ly$\alpha$,
Ly$\beta$, CIV, [CIII] 1909, MgII, SiIV, H$\alpha$, H$\beta$ and
H$\gamma$) as well as the blue bump centred at 3000 $\rm \AA$ are
also included in the QSO template SEDs. The Ly$\alpha$ forest has
been modelled according to Madau (1995). No reddening has been applied
to the QSO template SEDs. 

Using the spectroscopic data available for our sample, we attempt to
fine-tune the spectral line properties (i.e. equivalent widths and
relative intensities) of the QSO template SEDs. This is in order to
calibrate the HMP00 method on X-ray selected samples. We conclude
that the spectral line profiles used by HMP00 are also optimal 
for our X-ray selected AGN sample.

The template SED library of the HMP00 technique also comprises stellar
spectra for a range of spectral types including white dwarfs (Pickles
1998). These are used to identify candidate stars (Hatziminaoglou et al. 2002) 
within our X-ray selected sample. The observed
mean spectra of four different galaxy types (E/S0, Sbc, Scd,
Im) from  Coleman, Wu  \& Weedman (1980; CWW) are also used in this paper to 
estimate photometric redshifts for X-ray sources dominated by light
from the host galaxy.

For the photometric redshift estimation we employ the SDSS
5-band photometry using PSF magnitudes ({\sf psfMag} SDSS parameter)
for optically unresolved sources and petrosian magnitudes ({\sf
petroMag} SDSS parameter) for optically extended systems.


\begin{figure}

\setlength{\unitlength}{1mm}
\begin{picture}(80,90)
\includegraphics{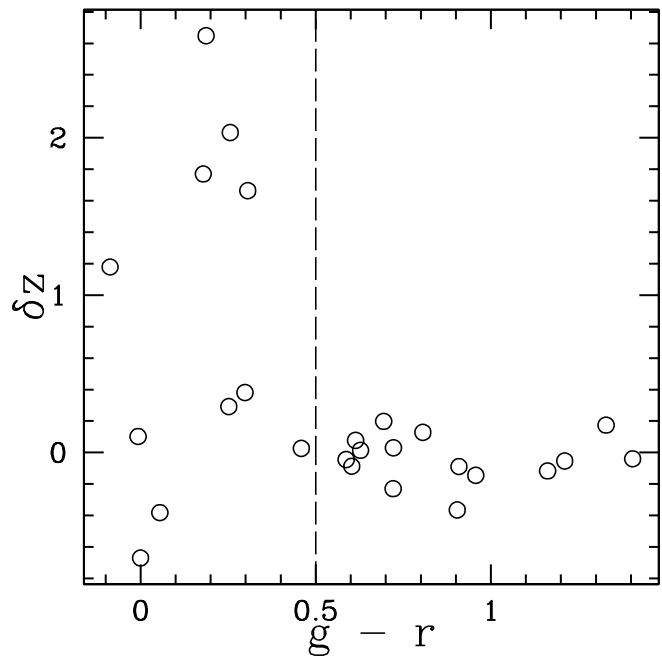}
\end{picture}

\caption{$\rm \delta z$ against $g-r$ colour for optically extended X-ray
sources spectroscopic
redshift measurements. Our method provides reasonable photometric redshift estimates
for relatively red sources ($g - r>0.5$). For bluer optical colours
the accuracy of our method is poorer most likely due to contamination
of the optical  light from the central engine.}
\label{fig:two}
\end{figure}

\section{Results}

\subsection{The spectroscopic sub-sample}
We apply the method described in the section above to the sub-sample
of X-ray sources with spectroscopic data available to assess the
success rate of our technique. The spectroscopic sub-sample comprises
71 optically unresolved  (e.g. point-like morphology) systems and 26 sources
with extended optical light profile. The morphological classification
is obtained from the SDSS. This is reliable at the 95 per cent
confidence level at the magnitude limit $r=21$\,mag (York et al. 2000;
Stoughton et  al. 2002). At fainter magnitudes the star/galaxy
classification becomes less robust.

We treat  extended and point-like sources separately using galaxy and
QSO templates respectively. X-ray sources with unresolved optical
light profile are most likely associated with distant powerful AGNs
where the light from the central engine dominates the optical
emission. QSO template SEDs are clearly more appropriate for these 
sources. The X-ray sources with extended optical light profile are
most likely  relatively nearby galaxies where the central source does
not dominate their optical light. Therefore, galaxy templates are
likely appropriate for these systems.

Before applying the HMP00 method to the full spectroscopic catalogue we
identify and exclude from our analysis Galactic star candidates. In
particular, we firstly identify optically unresolved (e.g. point-like
morphology) sources with low X-ray--to--optical flux ratio ($\log f_X /
f_{opt} < -1$; total of 13), i.e. in the region of the parameter space
occupied by Galactic stars (Stocke et al. 1991). Secondly, for these $\log
f_X /f_{opt} < -1$ sources we apply the HMP00 technique using both QSO and
stellar templates in an attempt to identify objects with optical SEDs
consistent with those of stars. This is because the $\log f_X / f_{opt}
<-1$ regime also comprises a small fraction of QSOs in addition to other
classes of sources. Of the 13 sources with $\log f_X /f_{opt} < -1$ a
total of 9 are best fit by stellar templates. All of them are
spectroscopically identified Galactic stars (Georgakakis et al. 2004). 
The remaining
4 sources are best fit by QSO templates. This classification is consistent
with the optical spectroscopy available for these objects (Georgakakis et
al. 2004). Therefore, our method based on both X-ray and optical broad
band information is successful in identifying all the Galactic stars
within the spectroscopic subsample. In the remaining of this paper we shall 
not consider the 9 galactic stars identified in our spectroscopic subsample.
  
For the remaining 62 optically unresolved X-ray sources
 we have applied the photometric redshift
technique of HMP00 using QSO templates only. The results are presented
in Table 1 and are compared with the spectroscopic redshift
measurements in Fig. 1. A total of 42 out of 62 objects ($\approx68$
per cent) have photometric redshift estimates correct within $\rm 
\delta z \la 0.3$. For this sample we estimate a fractional mean error
defined as

$$
\rm \overline{\delta z/(1+z)} = \sum     \left( 
                     \frac{z_{phot}-z_{spec}}{1+z_{spec}}
                     \right) /N,
$$

\noindent $\rm \overline{\delta z/(1+z)}=-0.02$, and an rms scatter defined as

$$
\rm \sigma_z^2 = \sum   \left(
                     \frac{z_{phot}-z_{spec}}{1+z_{spec}}
                     \right) ^2 /N,
$$

\noindent
$\rm \sigma_z=0.30$. The agreement between  photometric and spectroscopic
redshifts is fair out to z $\approx1.5$ and becomes poorer at higher
redshifts. As will be discussed in the next section, these problematic
redshift estimates are likely due to strong emission lines falling
between the SDSS filters.  As already discussed above, the star-galaxy
separation is 95 per cent reliable at $r=21$\,mag and degrades at
fainter magnitudes. This suggests that extended sources with
$r>21$\,mag may be erroneously classified point-like and vice versa
resulting in the use of wrong templates to determine their
redshift. In the optically point-like subsample there are 13 sources
with $r>21$\,mag (marked with a large circle in Fig. 1). As can be seen 
in Fig. 1, about half of them have
$\rm \delta z>0.3$. Using galaxy instead of QSO templates does
not improve their photometric redshift estimates. The low success rate
for these systems is most likely due to larger photometric errors
compared to optically brighter sources. 

For the 26 optically {\it extended} X-ray sources in the XMM-{\it
Newton}/2dF sample with spectroscopic data we have applied the HMP00
technique using {\it galaxy} SED templates only. The photometric
redshift estimates for these sources are also presented in Table 1. Fig.
2 plots $\rm \delta z$ as a function of $g-r$ colour for the
spectroscopic sub-sample of extended sources (total of 26 sources), while 
Fig. 3 plots spectroscopic against photometric redshift for the same sources. 
It is clear from these
figures that for sources with red optical colours ($g-r>0.5$\,mag) our
method performs well with a mean error of $\rm \overline{\delta z/(1+z)} =
-0.01$ and rms scatter $\rm \sigma_z=0.10$. For  objects with blue colours
($g-r<0.5$\,mag) however, our method is not as successful giving  
$\rm \overline{\delta z/(1+z)} = +0.59$ and $\rm \sigma_z=0.99$. The
photometric  redshifts of this blue X-ray source population cannot be
improved using QSO or starburst templates suggesting that their
optical light is a mix of both AGN and stellar emission. 

We further explore this possibility in an attempt to improve the
photometric redshift estimates for these sources by linearly combining
QSO (total of 3; see section 3) and galaxy (total of 4 from CWW)
templates. These are normalised at 5000\AA, although the choice of
the normalisation wavelength does not alter our results. For each of 
the 12 QSO/galaxy
SED pairs (3 QSO times 4 galaxy SEDs) we vary the normalisation of the
QSO component relative to the galaxy one between 0.1 and 0.9 in steps
of 0.2, and we then sum the weighted SED pairs. This results in 5 
linear combinations for each QSO/galaxy SED
pair and therefore a total of 60 combined QSO/galaxy templates. We
find that using these 60 templates does not significantly improve the
photometric redshift estimates of optically extended blue ($g-r<0.5$)
objects: only two additional sources are assigned photometric
redshifts accurate within $\rm \delta z \la 0.15$.

Finally, for the optically extended spectroscopic sub-sample there is
only 1 source  with $r>21$\,mag and therefore, we cannot comment on
the success rate of our method for fainter sources where the
star/galaxy separation becomes less reliable. For this
one source (\#10 in Table 1) the photometric redshift estimate is not 
in good agreement with the spectroscopic redshift measurement, 
but this most likely due to the blue colour of the object. 

\subsection{The spectroscopically unidentified sub-sample}
In this section we apply the HMP00 method to X-ray sources without
spectroscopic data available. This sub-sample comprises 96 sources of
which 45 have point-like optical morphology and 51 are optically
extended. From the point-like sub-sample we exclude one
source with $ \log ( f_{X} / f_{opt } ) < -1$ that is also best fit 
with stellar template. Based on the discussion in \S 4.1, 
this is likely to be a 
Galactic star. The results for the remaining 44 sources are presented
in Table 2. We also caution the reader that  there are  24 point-like
X-ray sources with  $r>21$\,mag for which, as discussed above, our
method is expected to perform less well.

The results for the 51 optically extended sources in the spectroscopically 
unidentified sub-sample are also presented
in Table 2. A total of 29 of these sources have red optical colours
($g-r>0.5$\,mag) for which our method gives acceptable photometric
redshift estimates. The remaining 22 sources however, have
$g-r<0.5$\,mag and as discussed above their photometric redshifts
are not reliable. Nevertheless, for the sake of completeness these
are also listed in Table 2.

\begin{figure}
\setlength{\unitlength}{1mm}
\begin{picture}(80,90)
\includegraphics{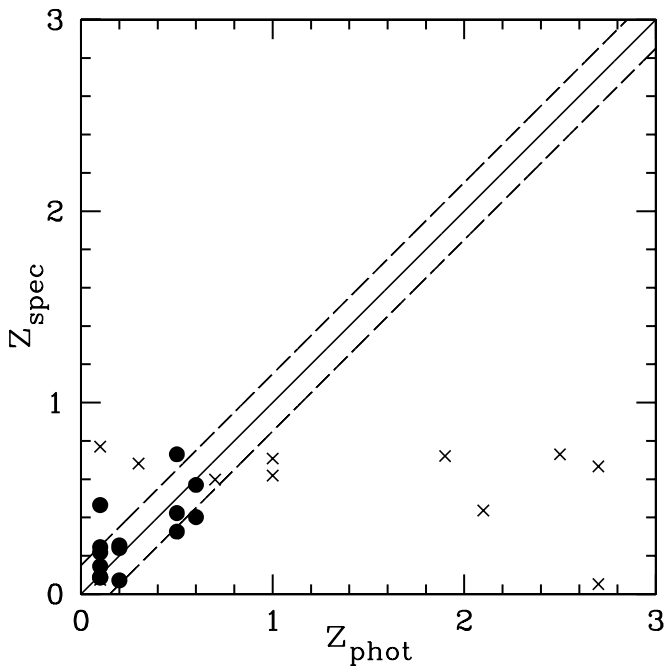}
\end{picture}

\caption{Comparison between z$_{spec}$ and z$_{phot}$ for X-ray  
sources with extended optical light profile and spectroscopic
redshift measurements. Filled circles are for red ($g-r>0.5$) and crosses 
are for blue ($g-r<0.5$) sources. The solid line is the z$_{spec}$ = z$_{phot}$
locus. The dashed lines define the $\rm \delta z = \pm 0.15$ envelope around
the  z$_{spec}$ = z$_{phot}$ line.}  
\label{fig:three}
\end{figure}



\begin{figure}

\setlength{\unitlength}{1mm}
\begin{picture}(80,90)
\includegraphics{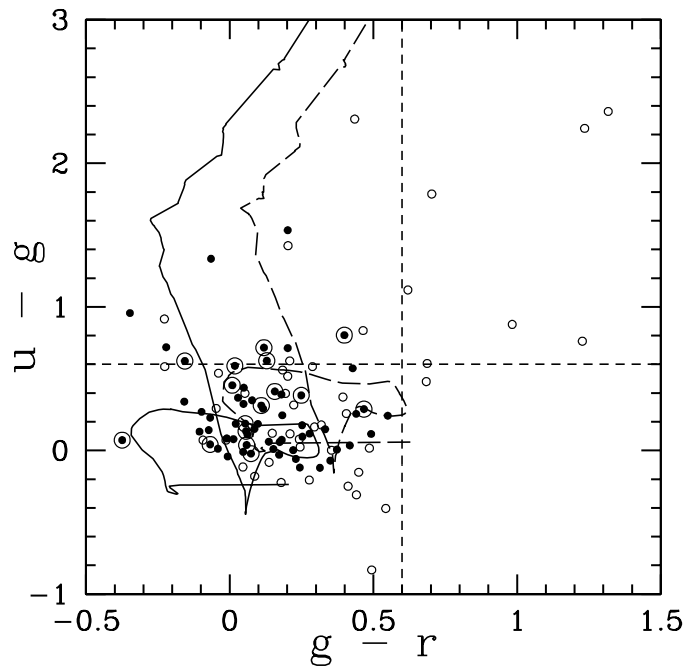}
\end{picture}

\caption{Colour-colour diagram of the X-ray sources with point-like
optical morphology. Open circles are for sources without z$_{spec}$,
filled circles correspond to sources in the  spectroscopic 
sub-sample. A large circle ontop of a symbol is for sources with 
problematic photometric redshift estimates ($\rm \delta z > 0.3$).
The two curves are colour-colour tracks of the HMP00 QSO
templates in the redshift range 0--6 for two different values of the
optical power-law spectral index, 0 and 1 for the solid and the
long-dashed lines respectively. For $u-g \protect \ga 0.6$ these templates
correspond to z $\protect \ga 2$. The short-dashed horizontal and 
vertical lines
correspond to $u-g=0.6$ and $g-r=0.6$. A total of 7 point-like X-ray
sources have colours redder than these limits. The photometric
redshifts of these sources are likely to be problematic since they
may have SEDs substantially different to those expected for 
optically selected QSOs.}  
\label{fig:four}
\end{figure}

\section{Discussion}

In this paper we describe a recipe for estimating photometric
redshifts for bright X-ray selected samples comprising both powerful 
distant QSOs and less luminous lower-z Seyfert type systems. 
Firstly, X-ray sources associated with Galactic stars are excluded
from our analysis by identifying optically unresolved objects with low
X-ray--to--optical flux ratio  ($\log  f_X/f_{opt}<-1$) and SEDs that 
are best fit with stellar templates. 
 The remaining X-ray sources are
split into optically point-like and extended systems. The
former are most likely associated with powerful QSOs where the central
engine dominates the optical light. The latter are low luminosity AGNs at 
 moderate redshifts and therefore their optical
colours also have a significant stellar component from
the host galaxy. For optically point-like objects we use QSO templates
to estimate their photometric redshifts while for optically extended
X-ray sources we use galactic SEDs. 

This method is applied to all the spectroscopically identified  X-ray sources
detected in the present XMM-{\it Newton}/2dF sample. For optically
extended sources with relatively red colours ($g-r>0.5$)
  the photometric redshift success rate is reasonably good: 
 i) $\sim$ 73 per cent have $\rm \delta z \leq 0.15$;
 ii) the fraction of non-catastrophic redshifts, i.e. with $\rm \delta z /(1+z) \leq 0.2 $,
 is 93 per cent; iii) the rms scatter is $\rm \sigma_z=0.10$. 
 The fraction of problematic photometric redshift estimates increases
however, for optically extended objects with $g-r<0.5$, most likely
due to contamination of their optical light by emission from the
central engine. For these sources we find that 
 27 per cent have $\rm \delta z /(1+z) \leq 0.2 $ and we estimate an 
 rms scatter of $\rm \sigma_z= 0.99$. We find no improvement when 
 QSO or star-forming galaxy templates, or linear combinations 
 of QSO and galaxy SEDs, are applied to these systems.
 In addition the application of a distance prior (z $<1$) 
 results in no significant improvement of the results with the rms scatter 
 being still large, i.e. $\rm \sigma_z=0.25$.  
 
For point-like sources we estimate an accuracy of $\rm \delta
z\la0.3$ for $\approx 68$ per cent of the sample. 
 About 75 per cent of these sources have $\rm \delta z /(1+z) \la 0.2 $. 
Fig. 1 shows that
there are degeneracies that affect the efficiency of our method,
particularly for  sources with z $\ga 1.5$. As discussed by Richards
et al. (2001) this is likely due to strong emission lines falling
between the SDSS filters at certain redshift ranges. Indeed, these 
authors use a different
method to that described here to estimate photometric redshifts for
optically selected SDSS QSOs. They find that photometric redshifts 
for 70 per cent of the sources in their
sample are correct within $\rm \delta z < 0.2$ with an rms scatter of
0.67, estimated on their whole sample ($\sim$2600 sources). Their success 
rate for 
{\it optically} selected QSOs is better than the one obtained here for {\it
X-ray} selected QSOs: 52 per cent have $\rm \delta z < 0.2$ and the rms
scatter is estimated to be 0.72 (based on our sample of 62 
sources).\footnote{We note that our {\it X-ray} selected spectroscopic 
sub-sample does not identify sources at redshifts larger than 3, a
redshift range where Richards et al. (2001; see their Fig. 2) have
an increased success rate. We also expect our method to perform 
similarly well at this redshift range (HMP00).} Nevertheless, 
similarly to our results, the accuracy
of their method is also lower for sources in the range $1.5 \la$ z
$\la 2.5$.  They attribute these problems to the MgII, CIV and
Ly$\alpha$ lines falling between the $u$ and $g$ bands at different redshifts.

We also explore the possibility that the colours of our X-ray selected  
AGNs are responsible for the problematic redshifts at
z $\ga 1.5$. Richards et al. (2001) found that about 4 per
cent of their optically selected AGN sample has colours redder than
those of typical QSOs at z $\la2.2$ (i.e. $u-g \ga 0.6$) suggesting either
reddening or high-z systems. A large fraction of these redder sources
have inaccurate photometric redshift estimates. Since X-ray selection
is least biased to obscured AGNs we also expect a non-negligible
fraction of QSOs with red colours  most likely due to dust
extinction. Estimating the redshift of these  systems is problematic
since their SEDs are likely to be very different to the templates used
by the HMP00 method.    

Fig. \ref{fig:four} plots $u-g$ against $g-r$ for all the X-ray
selected sources with unresolved optical morphology.  The curves are
the colour-colour tracks of the HMP00 QSO templates in the redshift
range 0--6 for two different values of the optical power-law spectral
index, 0 and 1 (solid and long-dashed line respectively). The z $\ga
2$ regime for  these models lies at $u-g \ga 0.6$. A number of sources
are redder than  $u-g > 0.6$ (total of 21) but these are most likely
high-z rather than dusty QSOs. Indeed, all sources with spectroscopic
identification and  $u-g \ga 0.6$ have z$_{spec} \ga
2$. Nevertheless, a striking result in Fig. \ref{fig:four} is the
non-negligible fraction of optically point-like X-ray sources that are  
red in both the  $u-g$ and the $g-r$ colours (7 per cent with $u-g > 
0.6$ and $g - r > 0.6$). These red colours cannot be accounted for by
the HMP00 QSO templates at any redshift and  suggest the presence
of dust.  Unfortunately spectroscopic redshifts are not available for
these sources to explore their nature. It is likely however, that the
photometric redshifts for these systems are problematic since their
colours are much redder than those of optically selected QSOs at any
redshift. A different set of templates SEDs are probably required for
such sources.


Babbedge et al. (2004) recently  
presented a method similar to the one described here to 
estimate photometric redshifts for both QSOs and galaxies in the 
ELAIS N1 and N2 fields as well as in  the {\it Chandra} Deep Field 
North region. 
 The additional criteria applied by our method are the use
 of the X-ray to optical flux ratios to identify stars
 and the separation between blue and red galaxies. 
 The application of these is essential in 
 bright X-ray samples which comprise an elevated 
 fraction  of both stars and  sources having blue colours.     
 The accuracy of the photometric redshift estimates
 of Babbedge et al., 
 in the case of the CDFN region,   
 is in reasonable agreement with those obtained here.
 The above authors find that 83 per cent of their QSO sample 
 have $\rm \delta z /(1+z)<0.2 $; in our case 
 this fraction amounts to 75 per cent. For the galaxies
 Babbedge et al. estimate that a fraction of 
 90 per cent has $\rm \delta z /(1+z)<0.2 $, while 
 we find a fraction of 93 per cent for the {\it red}
 galaxies only. We find however that there are limitations in 
 the application of photometric redshifts to the blue optically
 resolved sources.

%
%

\section{Conclusions}

 In this paper we are exploring the use of photometric redshifts 
 to bright X-ray samples. In particular, 
 we estimate photometric redshifts for a sample of 193
X-ray sources with optical counterparts in the XMM-{\it Newton}/2dF 
survey. This is a bright X-ray sample [$f_X(\rm 0.5-8 \,keV ) \approx
10^{-14}\, erg \, s^{-1} \, cm^{-2}$] consisting of both powerful
distant QSOs  and lower luminosity nearby AGNs. We split 
our sample into extended  and point-like  sources based on 
their optical light profile. We then employ the HMP00 technique to
determine photometric redshifts, using galaxy and QSO templates for
the extended and the point-like sources respectively.  A number of
stars can be immediately  identified  on the basis of their unresolved 
optical light profile, their low X-ray to optical flux ratio and their SEDs 
that are best fit with stellar templates. 

Comparison between photometric and spectroscopic redshifts for 62 
optically unresolved  sources
shows that for $\approx68$ per cent of them the photometric
redshift estimates are accurate within $\delta$z $\la0.3$.  The
photometric redshifts appear to be in good agreement with
spectroscopic ones up to z $\approx 1.5$. At larger redshifts 
the presence of emission lines in between filters limits the
efficiency of our method. We also find a non-negligible fraction (7\%)
of X-ray sources with point-like optical profile having colours redder
than those of optically selected QSOs. The photometric redshift
estimates for these sources are likely to be problematic.    

In the case of X-ray sources with  extended optical light profile, we
find that galaxy templates provide reasonable  
 (73 per cent have $\rm \delta z<0.15$)  photometric
redshifts at least for sources with red colours ($g-r > 0.5$). 
We find however that a combined galaxy-QSO template approach cannot 
 be successfully applied to optically extended sources with bluer
colours. These present poor photometric redshift estimates, 
 regardless on whether the estimates are based on galaxy or QSO templates, or any linear combination of the two.

\section*{Acknowledgments}

 We thank the anonymous referee for valuable comments and suggestions. 
This work is 
 funded by the European Union and the Greek Ministry
 of Development in the framework of the Programme 
 'Competitiveness-Promotion of Excellence in Technological Development and 
 Research-Action 3.3.1', Project 'X-ray Astrophysics with ESA's mission XMM',
 MIS-64564. The XMM-{\it
Newton}/2dF  survey data as well as part of the observations presented
here are electronically available at {\sf
http://www.astro.noa.gr/xray/}. 
  
The 2dF QSO Redshift Survey (2QZ) was compiled by the 2QZ survey team
from observations made with the 2-degree Field on the Anglo-Australian
Telescope. 

Funding for the creation and distribution of the SDSS Archive has been
provided by the Alfred P. Sloan Foundation, the Participating
Institutions, the National Aeronautics and Space Administration, the
National Science Foundation, the U.S. Department of Energy, the
Japanese Monbukagakusho, and the Max Planck Society. The SDSS Web site
is http://www.sdss.org/. The SDSS is managed by the Astrophysical
Research Consortium (ARC) for the Participating Institutions. The
Participating Institutions are The University of Chicago, Fermilab,
the Institute for Advanced Study, the Japan Participation Group, The
Johns Hopkins University, Los Alamos National Laboratory, the
Max-Planck-Institute for Astronomy (MPIA), the Max-Planck-Institute
for Astrophysics (MPA), New Mexico State University, University of
Pittsburgh, Princeton University, the United States Naval Observatory,
and the University of Washington.

\begin{table*}
\caption[]{Photometric redshifts for the 88 X-ray sources with spectroscopic identification in the XMM-{\it Newton}/2dF sample.}

\begin{tabular}{ccccrccccccc}

\hline \hline

\# & \multicolumn{3}{c}{$\alpha_{\rm opt}$ (J2000)} & 
\multicolumn{3}{c}{$\delta_{\rm opt}$ (J2000)} & z$_{phot}$ & z$_{spec}$ & r\footnote[5]{} & g-r\footnote[5]{} &  morph\footnote[5]{}\\

\hline

 & h & m & s & $^{\rm o}$ & $^{'}$ & $^{''}$ &            &  & & &        \\ 

\hline

1 & 13 & 40 & 41.43 & -00 & 17 & 26.52 & 0.740 &  2.754\footnote[2]{} & 20.108 &  0.129 &  P \\
2 & 13 & 40 & 49.98 &  00 & 26 & 44.84 & 2.120 &  1.496\footnote[4]{} & 21.368 &  0.057 &  P \\
3 & 13 & 40 & 50.18 &  00 & 07 & 28.09 & 0.940 &  0.901\footnote[4]{} & 19.761 &  0.183 &  P \\
4 & 13 & 40 & 59.07 &  00 & 05 & 04.63 & 0.600 &  2.599\footnote[1]{} & 18.885 &  0.018 &  P \\
5 & 13 & 40 & 59.24 & -00 & 19 & 44.94 & 1.600 &  1.857\footnote[1]{} & 19.956 &  0.086 &  P \\
6 & 13 & 41 & 01.35 & -00 & 27 & 32.15 & 1.000 &  0.619\footnote[4]{} & 20.024 &  0.298 &  E \\
7 & 13 & 41 & 02.92 &  00 & 15 & 47.20 & 1.330 &  1.038\footnote[4]{} & 19.628 &  0.253 &  P \\
8 & 13 & 41 & 03.22 & -00 & 24 & 48.71 & 0.640 &  1.565\footnote[4]{} & 20.644 &  0.054 &  P \\
9 & 13 & 41 & 07.90 & -00 & 17 & 25.37 & 2.200 &  0.798\footnote[4]{} & 21.324 &  0.110 &  P \\
10 & 13 & 41 & 16.96 &  00 & 21 & 50.40 & 1.900 &  0.721\footnote[4]{} & 21.172 & -0.087 &  E \\
11 & 13 & 41 & 17.89 & -00 & 23 & 21.55 & 0.500 &  0.423\footnote[4]{} & 19.639 &  0.614 &  E \\
12 & 13 & 41 & 20.81 &  00 & 14 & 50.60 & 2.200 &  2.550\footnote[4]{} & 21.794 &  0.029 &  P \\
13 & 13 & 41 & 21.33 & -00 & 13 & 52.28 & 0.620 &  0.734\footnote[2]{} & 19.249 & -0.098 &  P \\
14 & 13 & 41 & 22.88 & -00 & 22 & 45.80 & 2.130 &  1.919\footnote[2]{} & 19.372 & -0.041 &  P \\
15 & 13 & 41 & 27.19 &  00 & 14 & 13.81 & 1.600 &  1.698\footnote[2]{} & 19.457 &  0.180 &  P \\
16 & 13 & 41 & 27.92 &  00 & 32 & 11.65 & 0.120 &  1.764\footnote[2]{} & 21.719 & -0.374 &  P \\
17 & 13 & 41 & 33.38 & -00 & 24 & 32.11 & 0.200 &  0.072\footnote[1]{} & 16.272 &  0.806 &  E \\
18 & 13 & 41 & 33.67 & -00 & 27 & 03.96 & 1.000 &  1.341\footnote[2]{} & 19.231 &  0.252 &  P \\
19 & 13 & 41 & 35.63 & -00 & 00 & 59.11 & 1.500 &  1.602\footnote[1]{} & 19.239 &  0.098 &  P \\
20 & 13 & 41 & 35.69 & -00 & 14 & 04.45 & 1.460 &  1.342\footnote[4]{} & 20.387 &  0.550 &  P \\
21 & 13 & 41 & 37.69 & -00 & 25 & 55.31 & 2.700 &  0.052\footnote[4]{} & 16.373 &  0.187 &  E \\
22 & 13 & 41 & 40.25 &  00 & 15 & 42.55 & 0.200 &  0.254\footnote[4]{} & 18.597 &  1.211 &  E \\
23 & 13 & 41 & 42.83 &  00 & 12 & 38.88 & 0.490 &  0.790\footnote[2]{} & 19.832 & -0.073 &  P \\
24 & 13 & 41 & 47.77 &  00 & 31 & 07.14 & 1.320 &  1.335\footnote[4]{} & 20.563 &  0.314 &  P \\
25 & 13 & 41 & 55.72 & -00 & 22 & 32.56 & 0.700 &  2.426\footnote[2]{} & 20.071 &  0.119 &  P \\
26 & 13 & 41 & 56.86 &  00 & 30 & 10.33 & 1.420 &  1.238\footnote[1]{} & 18.966 &  0.221 &  P \\
27 & 13 & 42 & 11.95 &  00 & 29 & 50.93 & 0.600 &  0.570\footnote[2]{} & 20.418 &  0.722 &  E \\
28 & 13 & 42 & 12.20 & -00 & 17 & 37.72 & 0.100 &  0.086\footnote[1]{} & 16.547 &  0.628 &  E \\
29 & 13 & 42 & 14.40 & -00 & 28 & 43.43 & 0.600 &  0.402\footnote[4]{} & 19.766 &  0.694 &  E \\
30 & 13 & 42 & 17.20 &  00 & 35 & 30.30 & 1.320 &  1.300\footnote[4]{} & 20.515 &  0.418 &  P \\
31 & 13 & 42 & 19.12 &  00 & 02 & 54.74 & 1.310 &  1.185\footnote[2]{} & 20.322 &  0.440 &  P \\
32 & 13 & 42 & 27.95 & -00 & 07 & 13.33 & 1.360 &  1.233\footnote[4]{} & 19.363 &  0.278 &  P \\
33 & 13 & 42 & 28.07 &  00 & 30 & 18.86 & 0.700 &  0.598\footnote[4]{} & 20.838 & -0.007 &  E \\
34 & 13 & 42 & 32.38 & -00 & 31 & 51.06 & 1.040 &  1.209\footnote[1]{} & 18.560 &  0.428 &  P \\
35 & 13 & 42 & 32.95 & -00 & 15 & 50.69 & 0.830 &  2.132\footnote[2]{} & 19.686 &  0.249 &  P \\
36 & 13 & 42 & 33.71 & -00 & 11 & 48.16 & 0.510 &  0.516\footnote[1]{} & 19.066 & -0.158 &  P \\
37 & 13 & 42 & 46.27 & -00 & 35 & 43.76 & 0.820 &  0.787\footnote[1]{} & 18.141 & -0.105 &  P \\
38 & 13 & 42 & 53.03 & -00 & 18 & 12.38 & 3.100 &  2.920\footnote[3]{} & 21.257 &  0.202 &  P \\
39 & 13 & 42 & 55.43 &  00 & 06 & 34.78 & 2.100 &  0.436\footnote[2]{} & 19.334 &  0.306 &  E \\
40 & 13 & 42 & 56.52 &  00 & 00 & 57.24 & 0.800 &  0.804\footnote[1]{} & 18.779 &  0.048 &  P \\
41 & 13 & 43 & 01.58 & -00 & 29 & 50.68 & 2.130 &  2.062\footnote[2]{} & 19.980 &  0.013 &  P \\
42 & 13 & 43 & 07.13 & -00 & 27 & 51.26 & 0.100 &  0.770\footnote[4]{} & 20.849 &  0.000 &  E \\
43 & 13 & 43 & 07.77 &  00 & 27 & 20.56 & 2.500 &  2.310\footnote[4]{} & 20.874 & -0.347 &  P \\
44 & 13 & 43 & 09.20 & -00 & 22 & 56.46 & 1.410 &  2.210\footnote[3]{} & 21.200 &  0.467 &  P \\
45 & 13 & 43 & 13.37 & -00 & 05 & 01.32 & 2.700 &  0.667\footnote[3]{} & 20.456 &  0.256 &  E \\
46 & 13 & 43 & 14.42 &  00 & 06 & 47.38 & 1.440 &  1.376\footnote[4]{} & 20.243 &  0.244 &  P \\
47 & 13 & 43 & 14.88 &  00 & 25 & 29.10 & 1.500 &  1.468\footnote[2]{} & 19.371 &  0.174 &  P \\
48 & 13 & 43 & 23.66 &  00 & 12 & 23.33 & 0.840 &  0.874\footnote[1]{} & 18.319 &  0.078 &  P \\
49 & 13 & 43 & 24.20 & -00 & 07 & 29.64 & 1.600 &  1.910\footnote[3]{} & 22.003 &  0.202 &  P \\
50 & 13 & 43 & 24.29 & -00 & 20 & 29.76 & 1.630 &  1.890\footnote[2]{} & 19.811 & -0.007 &  P \\
51 & 13 & 43 & 29.22 &  00 & 01 & 32.77 & 1.600 &  2.350\footnote[3]{} & 21.043 &  0.399 &  P \\
52 & 13 & 43 & 31.44 &  00 & 11 & 08.56 & 1.430 &  1.280\footnote[2]{} & 19.374 &  0.181 &  P \\
53 & 13 & 43 & 32.68 & -00 & 02 & 03.23 & 0.500 &  1.600\footnote[4]{} & 20.076 &  0.009 &  P \\
54 & 13 & 43 & 39.68 &  00 & 29 & 37.79 & 2.300 &  2.378\footnote[2]{} & 20.291 & -0.221 &  P \\
55 & 13 & 43 & 42.53 &  00 & 18 & 03.10 & 1.400 &  1.115\footnote[4]{} & 20.215 &  0.172 &  P \\
56 & 13 & 43 & 47.45 &  00 & 20 & 23.78 & 0.200 &  0.240\footnote[4]{} & 18.147 &  1.405 &  E \\
57 & 13 & 43 & 47.58 & -00 & 23 & 36.38 & 1.060 &  1.057\footnote[2]{} & 19.651 &  0.332 &  P \\
58 & 13 & 43 & 51.06 &  00 & 04 & 34.72 & 0.100 &  0.074\footnote[1]{} & 16.753 &  0.459 &  E \\

\hline
\end{tabular}
\label{tab1}
\end{table*}

\setcounter{table}{0}

\begin{table*}
\caption{- {\it continued}.}
\begin{tabular}{ccccrccccccc}

\hline \hline

\# & \multicolumn{3}{c}{$\alpha_{\rm opt}$ (J2000)} & 
\multicolumn{3}{c}{$\delta_{\rm opt}$ (J2000)} & z$_{phot}$ & z$_{spec}$ & r\footnote[5]{} & g-r\footnote[5]{} &  morph\footnote[5]{}\\

\hline

 & h & m & s & $^{\rm o}$ & $^{'}$ & $^{''}$ &            &  & & &        \\ 

\hline

59 & 13 & 43 & 53.48 & -00 & 05 & 20.08 & 1.430 &  1.139\footnote[2]{} & 20.338 &  0.230 &  P \\
60 & 13 & 43 & 59.90 &  00 & 00 & 46.04 & 1.470 &  1.410\footnote[3]{} & 19.868 &  0.152 &  P \\
61 & 13 & 44 & 01.95 &  00 & 00 & 03.46 & 0.100 &  0.245\footnote[4]{} & 19.095 &  0.957 &  E \\
62 & 13 & 44 & 12.89 & -00 & 30 & 05.47 & 1.000 &  0.708\footnote[1]{} & 20.298 &  0.252 &  E \\
63 & 13 & 44 & 14.08 & -00 & 29 & 50.10 & 2.200 &  0.538\footnote[2]{} & 19.383 &  0.157 &  P \\
64 & 13 & 44 & 19.15 & -00 & 32 & 56.29 & 0.250 &  1.729\footnote[4]{} & 21.112 &  0.059 &  P \\
65 & 13 & 44 & 20.08 & -00 & 31 & 10.52 & 0.300 &  0.682\footnote[2]{} & 20.335 &  0.055 &  E \\
66 & 13 & 44 & 20.82 & -00 & 04 & 58.91 & 0.500 &  0.326\footnote[4]{} & 19.872 &  1.329 &  E \\
67 & 13 & 44 & 20.89 &  00 & 02 & 27.02 & 2.110 &  1.887\footnote[2]{} & 20.177 & -0.010 &  P \\
68 & 13 & 44 & 22.17 & -00 & 34 & 19.38 & 0.100 &  0.217\footnote[4]{} & 18.201 &  1.162 &  E \\
69 & 13 & 44 & 23.95 & -00 & 28 & 46.42 & 0.500 &  2.235\footnote[4]{} & 21.567 & -0.156 &  P \\
70 & 13 & 44 & 24.55 & -00 & 24 & 12.74 & 1.600 &  1.767\footnote[3]{} & 21.072 & -0.068 &  P \\
71 & 13 & 44 & 24.57 & -00 & 13 & 03.14 & 1.340 &  1.110\footnote[3]{} & 20.414 &  0.373 &  P \\
72 & 13 & 44 & 24.57 & -00 & 06 & 17.03 & 2.120 &  2.003\footnote[2]{} & 20.258 &  0.062 &  P \\
73 & 13 & 44 & 25.22 & -00 & 18 & 26.93 & 2.130 &  1.970\footnote[3]{} & 21.718 &  0.047 &  P \\
74 & 13 & 44 & 25.95 & -00 & 00 & 56.20 & 1.410 &  1.097\footnote[1]{} & 18.504 &  0.136 &  P \\
75 & 13 & 44 & 27.91 & -00 & 30 & 28.69 & 1.330 &  1.374\footnote[2]{} & 18.533 &  0.492 &  P \\
76 & 13 & 44 & 36.15 &  00 & 33 & 23.76 & 1.700 &  1.430\footnote[4]{} & 20.160 &  0.350 &  P \\
77 & 13 & 44 & 37.02 &  00 & 30 & 55.30 & 1.600 &  1.519\footnote[2]{} & 19.648 &  0.021 &  P \\
78 & 13 & 44 & 37.41 & -00 & 32 & 37.36 & 2.400 &  2.520\footnote[4]{} & 21.841 & -0.065 &  P \\
79 & 13 & 44 & 52.91 &  00 & 05 & 20.26 & 0.100 &  0.087\footnote[1]{} & 16.325 &  0.603 &  E \\
80 & 13 & 44 & 54.64 & -00 & 19 & 07.57 & 0.900 &  0.852\footnote[2]{} & 20.397 &  0.116 &  P \\
81 & 13 & 44 & 55.82 & -00 & 01 & 16.46 & 2.200 &  2.198\footnote[2]{} & 20.893 &  0.049 &  P \\
82 & 13 & 44 & 58.26 &  00 & 16 & 22.87 & 0.100 &  0.145\footnote[4]{} & 18.022 &  0.587 &  E \\
83 & 13 & 44 & 58.26 & -00 & 35 & 57.59 & 0.100 &  0.465\footnote[4]{} & 19.389 &  0.904 &  E \\
84 & 13 & 44 & 59.45 & -00 & 15 & 59.54 & 1.470 &  0.245\footnote[1]{} & 17.529 &  0.074 &  P \\
85 & 13 & 45 & 08.04 & -00 & 05 & 26.70 & 2.500 &  0.730\footnote[4]{} & 19.776 &  0.179 &  E \\
86 & 13 & 45 & 12.30 & -00 & 31 & 31.08 & 2.120 &  1.690\footnote[2]{} & 20.521 & -0.068 &  P \\
87 & 13 & 45 & 15.31 &  00 & 15 & 22.54 & 0.100 &  0.089\footnote[1]{} & 16.123 &  0.909 &  E \\
88 & 13 & 45 & 19.09 &  00 & 31 & 00.44 & 0.500 &  0.730\footnote[4]{} & 19.511 &  0.721 &  E \\
 
\hline

\multicolumn{12}{l}{\footnote[1]{}spectroscopy from SDSS} \\
\multicolumn{12}{l}{\footnote[2]{}spectroscopy from 2QZ} \\
\multicolumn{12}{l}{\footnote[3]{}spectroscopy from AAT with AUTOFIB or LDSS} \\
\multicolumn{12}{l}{\hspace{3mm} (Georgantopoulos et al. 1996)} \\ 
\multicolumn{12}{l}{\footnote[4]{}our own spectroscopy (Georgakakis et
al. 2004)} \\ 
\multicolumn{12}{l}{\footnote[5]{}r: SDSS $r$ band magnitude; g-r: SDSS $g-r$ colour in mag; morph: optical} \\
\multicolumn{12}{l}{\hspace{3mm} morphology (P - point-like; E - extended)} \\

\hline

\end{tabular}
\label{tab2}
\end{table*}

\begin{table*}
\caption[]{Photometric redshifts for the 95 X-ray sources without spectroscopic identification in the XMM-{\it Newton}/2dF sample.} 
\begin{tabular}{ccccrcccccc}

\hline \hline

\# & \multicolumn{3}{c}{$\alpha_{\rm opt}$ (J2000)} & 
\multicolumn{3}{c}{$\delta_{\rm opt}$ (J2000)} & z$_{phot}$ & r\footnote[1]{} & g-r\footnote[1]{} &  morph\footnote[1]{}\\

\hline

 & h & m & s & $^{\rm o}$ & $^{'}$ & $^{''}$ &            &  & &        \\ 

\hline

1 & 13 & 40 & 38.61 &  00 & 19 & 19.52 & 0.500 & 19.626 &  0.770 &  E \\
2 & 13 & 40 & 43.05 &  00 & 17 & 03.70 & 3.100 & 22.048 &  0.154 &  E \\
3 & 13 & 40 & 44.02 &  00 & 27 & 03.96 & 4.280 & 20.253 &  1.235 &  P \\
4 & 13 & 40 & 44.89 &  00 & 26 & 18.13 & 2.140 & 21.990 & -0.093 &  P \\
5 & 13 & 40 & 44.99 & -00 & 24 & 03.92 & 0.500 & 20.185 &  1.510 &  E \\
6 & 13 & 40 & 52.49 &  00 & 20 & 06.76 & 0.800 & 21.524 &  0.459 &  E \\
7 & 13 & 40 & 58.59 & -00 & 22 & 33.38 & 3.300 & 18.319 &  0.703 &  P \\
8 & 13 & 40 & 59.90 & -00 & 34 & 08.04 & 1.600 & 21.914 &  0.209 &  P \\
9 & 13 & 41 & 14.44 &  00 & 20 & 31.02 & 3.000 & 21.168 &  0.982 &  E \\
10 & 13 & 41 & 18.35 &  00 & 26 & 41.10 & 1.700 & 21.782 &  0.087 &  P \\
11 & 13 & 41 & 22.41 &  00 & 28 & 25.93 & 2.300 & 21.903 & -0.227 &  P \\
12 & 13 & 41 & 23.78 &  00 & 37 & 32.12 & 0.200 & 22.199 &  0.102 &  E \\
13 & 13 & 41 & 25.26 & -00 & 28 & 09.41 & 2.900 & 22.704 & -0.269 &  E \\
14 & 13 & 41 & 27.15 &  00 & 23 & 28.10 & 2.200 & 21.389 &  0.184 &  P \\
15 & 13 & 41 & 27.22 &  00 & 18 & 02.88 & 2.800 & 21.621 &  0.026 &  E \\
16 & 13 & 41 & 27.28 &  00 & 20 & 25.66 & 0.400 & 22.721 & -0.039 &  P \\
17 & 13 & 41 & 27.38 & -00 & 16 & 59.77 & 1.900 & 21.559 &  0.441 &  P \\
18 & 13 & 41 & 28.33 & -00 & 31 & 20.39 & 0.600 & 20.637 &  0.908 &  E \\
19 & 13 & 41 & 28.91 &  00 & 29 & 55.54 & 0.500 & 20.797 & -0.047 &  P \\
20 & 13 & 41 & 29.42 & -00 & 25 & 41.70 & 0.500 & 20.715 &  0.899 &  E \\
21 & 13 & 41 & 29.48 &  00 & 29 & 09.96 & 0.800 & 21.047 &  0.879 &  E \\
22 & 13 & 41 & 31.01 &  00 & 16 & 14.02 & 2.200 & 21.510 &  0.288 &  P \\
23 & 13 & 41 & 31.60 & -00 & 01 & 08.22 & 0.200 & 21.846 &  0.571 &  E \\
24 & 13 & 41 & 34.41 &  00 & 28 & 08.69 & 0.700 & 20.266 &  0.925 &  E \\
25 & 13 & 41 & 44.98 & -00 & 26 & 20.98 & 0.600 & 21.035 &  0.787 &  E \\
26 & 13 & 41 & 50.96 &  00 & 25 & 41.63 & 2.600 & 20.716 &  0.173 &  E \\
27 & 13 & 41 & 52.81 &  00 & 33 & 07.34 & 1.500 & 21.926 &  0.449 &  P \\
28 & 13 & 41 & 53.77 & -00 & 18 & 07.45 & 2.000 & 22.073 &  0.277 &  P \\
29 & 13 & 42 & 01.82 &  00 & 21 & 32.62 & 0.400 & 19.045 &  1.395 &  E \\
30 & 13 & 42 & 13.43 & -00 & 23 & 26.56 & 0.600 & 21.730 &  0.186 &  E \\
31 & 13 & 42 & 13.91 &  00 & 22 & 12.36 & 1.600 & 21.764 &  0.684 &  P \\
32 & 13 & 42 & 17.44 &  00 & 32 & 41.60 & 2.010 & 19.778 &  0.179 &  P \\
33 & 13 & 42 & 34.32 &  00 & 12 & 39.46 & 2.100 & 20.279 &  0.137 &  P \\
34 & 13 & 42 & 34.44 &  00 & 32 & 25.51 & 1.570 & 20.984 &  0.354 &  P \\
35 & 13 & 42 & 37.82 & -00 & 15 & 22.68 & 1.500 & 20.330 &  0.464 &  P \\
36 & 13 & 42 & 40.88 &  00 & 16 & 48.04 & 0.500 & 20.255 &  1.471 &  E \\
37 & 13 & 42 & 48.33 &  00 & 26 & 01.90 & 3.000 & 21.747 &  0.435 &  P \\
38 & 13 & 42 & 48.89 & -00 & 05 & 49.20 & 1.640 & 21.842 &  0.046 &  P \\
39 & 13 & 42 & 49.52 & -00 & 12 & 48.38 & 4.420 & 20.417 &  1.317 &  P \\
40 & 13 & 42 & 55.31 &  00 & 05 & 53.16 & 0.700 & 22.194 &  0.395 &  E \\
41 & 13 & 42 & 59.41 & -00 & 25 & 01.24 & 0.700 & 20.801 &  0.676 &  E \\
42 & 13 & 43 & 01.34 &  00 & 26 & 34.08 & 1.500 & 20.424 &  0.687 &  P \\
43 & 13 & 43 & 02.21 & -00 & 13 & 04.73 & 1.200 & 15.658 &  1.433 &  E \\
44 & 13 & 43 & 05.33 & -00 & 19 & 26.44 & 0.700 & 21.210 &  0.671 &  E \\
45 & 13 & 43 & 10.75 &  00 & 33 & 02.81 & 0.400 & 20.827 &  0.284 &  E \\
46 & 13 & 43 & 13.66 & -00 & 26 & 19.61 & 0.010 & 22.423 &  0.494 &  P \\
47 & 13 & 43 & 16.58 &  00 & 37 & 21.90 & 1.430 & 19.668 &  0.295 &  P \\
48 & 13 & 43 & 22.32 &  00 & 09 & 51.70 & 1.300 & 21.325 &  0.543 &  P \\
49 & 13 & 43 & 22.93 &  00 & 24 & 18.04 & 0.700 & 21.939 &  0.488 &  E \\
50 & 13 & 43 & 27.00 &  00 & 24 & 32.15 & 1.200 & 21.133 &  0.319 &  P \\
51 & 13 & 43 & 27.10 & -00 & 16 & 14.84 & 0.900 & 22.868 & -0.463 &  E \\
52 & 13 & 43 & 31.47 &  00 & 24 & 51.59 & 1.150 & 20.272 &  0.245 &  P \\
53 & 13 & 43 & 42.39 &  00 & 06 & 44.35 & 0.300 & 20.128 &  1.035 &  E \\
54 & 13 & 43 & 54.75 &  00 & 31 & 50.41 & 2.000 & 21.849 &  0.412 &  P \\
55 & 13 & 43 & 56.46 &  00 & 02 & 09.64 & 3.400 & 20.613 &  0.983 &  P \\
56 & 13 & 44 & 06.13 &  00 & 22 & 12.43 & 1.060 & 20.749 &  0.406 &  P \\
57 & 13 & 44 & 09.89 &  00 & 29 & 23.60 & 0.820 & 20.883 &  0.223 &  P \\
58 & 13 & 44 & 14.15 &  00 & 16 & 42.24 & 0.700 & 19.347 &  0.320 &  E \\

\hline
\end{tabular}
\label{tab3}
\end{table*}

\setcounter{table}{1}

\begin{table*}
\caption{- {\it continued}.}
\begin{tabular}{ccccrccccccc}

\hline \hline

\# & \multicolumn{3}{c}{$\alpha_{\rm opt}$ (J2000)} & 
\multicolumn{3}{c}{$\delta_{\rm opt}$ (J2000)} & z$_{phot}$ & r\footnote[1]{} & g-r\footnote[1]{} &  morph\footnote[1]{}\\

\hline

 & h & m & s & $^{\rm o}$ & $^{'}$ & $^{''}$ &            &  & &        \\ 

\hline

59 & 13 & 44 & 17.06 &  00 & 23 & 54.24 & 0.300 & 21.450 &  0.828 &  E \\
60 & 13 & 44 & 18.72 & -00 & 09 & 10.33 & 0.600 & 20.745 &  0.624 &  E \\
61 & 13 & 44 & 18.93 &  00 & 06 & 52.52 & 2.900 & 23.450 & -0.716 &  E \\
62 & 13 & 44 & 19.93 &  00 & 04 & 16.46 & 4.200 & 19.985 &  1.387 &  E \\
63 & 13 & 44 & 20.52 & -00 & 05 & 11.62 & 3.400 & 20.988 &  1.227 &  P \\
64 & 13 & 44 & 25.03 &  00 & 01 & 23.23 & 2.100 & 21.259 & -0.013 &  P \\
65 & 13 & 44 & 25.13 &  00 & 25 & 00.30 & 0.800 & 21.986 &  0.202 &  P \\
66 & 13 & 44 & 27.33 & -00 & 02 & 18.31 & 1.600 & 21.718 &  0.394 &  P \\
67 & 13 & 44 & 28.14 &  00 & 25 & 48.86 & 1.100 & 21.018 &  0.372 &  E \\
68 & 13 & 44 & 28.47 & -00 & 17 & 03.30 & 1.460 & 21.444 &  0.210 &  P \\
69 & 13 & 44 & 29.20 & -00 & 05 & 53.02 & 0.600 & 20.378 &  1.040 &  E \\
70 & 13 & 44 & 31.00 & -00 & 15 & 29.70 & 0.300 & 21.550 &  0.686 &  E \\
71 & 13 & 44 & 33.26 & -00 & 33 & 09.97 & 1.200 & 22.629 & -0.311 &  E \\
72 & 13 & 44 & 33.45 & -00 & 24 & 54.47 & 6.000 & 22.915 & -0.160 &  E \\
73 & 13 & 44 & 37.06 & -00 & 10 & 30.90 & 1.300 & 20.915 &  0.486 &  P \\
74 & 13 & 44 & 38.31 & -00 & 00 & 51.05 & 1.200 & 21.348 & -0.015 &  E \\
75 & 13 & 44 & 41.20 & -00 & 24 & 23.04 & 0.800 & 23.189 & -0.389 &  E \\
76 & 13 & 44 & 43.66 & -00 & 35 & 18.46 & 0.500 & 21.765 & -0.226 &  P \\
77 & 13 & 44 & 45.49 &  00 & 16 & 05.59 & 2.300 & 21.246 &  0.203 &  P \\
78 & 13 & 44 & 50.74 &  00 & 16 & 45.34 & 1.500 & 19.940 &  0.242 &  P \\
79 & 13 & 44 & 51.48 & -00 & 23 & 00.60 & 0.400 & 19.646 &  1.271 &  E \\
80 & 13 & 44 & 51.56 &  00 & 24 & 46.80 & 1.600 & 21.841 &  0.549 &  E \\
81 & 13 & 44 & 52.12 & -00 & 36 & 53.32 & 0.810 & 20.306 &  0.053 &  P \\
82 & 13 & 44 & 54.05 &  00 & 36 & 52.74 & 0.500 & 20.580 &  0.862 &  E \\
83 & 13 & 45 & 00.47 &  00 & 09 & 33.16 & 0.100 & 19.109 &  0.911 &  E \\
84 & 13 & 45 & 05.65 & -00 & 10 & 15.38 & 0.200 & 21.616 &  0.808 &  E \\
85 & 13 & 45 & 06.17 &  00 & 33 & 14.00 & 1.420 & 19.505 &  0.148 &  P \\
86 & 13 & 45 & 07.47 &  00 & 04 & 06.82 & 0.500 & 19.529 &  0.842 &  E \\
87 & 13 & 45 & 08.36 &  00 & 22 & 27.98 & 0.100 & 21.055 &  1.017 &  E \\
88 & 13 & 45 & 09.64 &  00 & 20 & 52.80 & 0.500 & 19.542 &  1.338 &  E \\
89 & 13 & 45 & 09.80 & -00 & 27 & 40.75 & 1.500 & 20.228 &  0.620 &  P \\
90 & 13 & 45 & 10.33 &  00 & 18 & 52.31 & 0.200 & 19.174 &  0.797 &  E \\
91 & 13 & 45 & 12.53 & -00 & 17 & 45.17 & 2.140 & 21.443 &  0.194 &  P \\
92 & 13 & 45 & 15.77 &  00 & 28 & 51.06 & 0.900 & 21.292 &  0.351 &  E \\
93 & 13 & 45 & 15.88 & -00 & 17 & 53.81 & 2.700 & 21.218 &  0.330 &  E \\
94 & 13 & 45 & 19.85 &  00 & 28 & 50.66 & 0.700 & 20.244 &  0.362 &  E \\
95 & 13 & 45 & 22.59 &  00 & 01 & 00.19 & 0.400 & 20.940 &  0.422 &  E \\

\hline 

\multicolumn{11}{l}{\footnote[1]{}r: SDSS $r$ band magnitude; g-r: SDSS $g-r$ colour in mag; morph: optical} \\
\multicolumn{11}{l}{\hspace{3mm} morphology (P - point-like; E - extended)} \\

\hline 
\end{tabular}
\label{tab4}
\end{table*}

\label{lastpage}


\begin{thebibliography}{}

\bibitem[]{1}
Babbedge T.S.R., et al., 2004, MNRAS, 353, 654
\bibitem[]{1a}
Barcons X., et al., 2002, A\&A, 382, 522

\bibitem[]{2}
Barger A., Cowie L.L., Brandt W.N., Capak P., Garmire G.P.,
Hornschemeier A.E., Steffen A.T., Wehner E.H., 2002, AJ, 124, 1839 

\bibitem[]{4}
Brandt W.N., et al., 2001, AJ, 122, 2810

\bibitem[]{5}
Brandt W.N., et al., 2002, ApJ, 569, L5

\bibitem[]{6}
Budavari T., et al., 2001, AJ, 122, 1163

\bibitem[]{7}
Budavari T., et al., 2003, ApJ, 595, 59

\bibitem[]{7a}
Coleman G. D., Wu C.-C., Weedman D. W., 1980, ApJ, 43, 393

\bibitem[]{8}
Croom S.M., Smith R.J., Boyle B.J., Shanks T., Loaring N.S., Miller,
L., Lewis I.J., 2001, MNRAS, 322, 29 

\bibitem[]{9}
Csabai I., et al., 2003, AJ, 125, 580

\bibitem[]{10}
Downes A.J.B., Peacock J.A., Savage A., Carrie D.R., 1986, MNRAS, 218, 31

\bibitem[]{12}
Fukugita M., Ichikawa T., Gunn J.E., Doi M., Shimasaku K., Schneider
D. P., 1996, AJ, 111, 1748 

\bibitem[]{13}
Georgakakis A., Georgantopoulos I., Stewart G.C., Shanks T., Boyle
B.J., 2003, MNRAS, 344, 161

\bibitem[]{14}
Georgakakis A., Georgantopoulos I., Vallbe M., Kolokotronis V.,
Basilakos S., Plionis M., Stewart G.C., Shanks T., Boyle B.J., 2004,
MNRAS, 349, 135

\bibitem[]{15}
Georgantopoulos I., Stewart G.C., Shanks T., Boyle B.J., Griffiths
R.E., 1996, MNRAS, 280, 276 

\bibitem[]{16}
Georgantopoulos I., Georgakakis A., Akylas A., Stewart G.C., Giannakis
O., Shanks T., Kitsionas S., 2004, MNRAS, 352, 91 

\bibitem[]{17}
Giacconi R., et al., 2001, ApJ, 551, 624

\bibitem[]{18}
Gonzalez A.H., Maccarone T.J., 2002, ApJ, 581, 155

\bibitem[]{19}
Green P.J., et al., 2004, ApJS, 150, 43

\bibitem[]{20}
Hasinger G., et al., 2001, A\&A, 365, L45

\bibitem[]{21}
Hatziminaoglou E., Mathez G., Pell\'{o} R., 2000, A\&A, 359, 9

\bibitem[]{22}
Hatziminaoglou E., et al., 2002, A\&A, 389, 81

\bibitem[]{23}
Kim D.-W., et al., 2004a, ApJS, 150, 19

\bibitem[]{24}
Kim D.-W., et al.,  2004b, ApJ, 600, 59

\bibitem[]{25}
Madau P., 1995, ApJ, 441, 18

\bibitem[]{26}
Mainieri V., Bergeron J., Hasinger G., Lehmann I., Rosati P., Schmidt
M., Szokoly G., Della Ceca R., 2002, A\&A, 393, 425  

\bibitem[]{27}
Mobasher B., et al.,  2004, ApJ, 600, L167

\bibitem[]{28}
Pickles A.J., 1998, MNRAS, 110, 863

\bibitem[]{29a}
Richards G.T., et al., 2001, AJ, 122, 1151

\bibitem[]{29b}
Richards G.T., et al., 2002, AJ, 123, 2945

\bibitem[]{30}
Stocke J.T., Morris S.L., Gioia I.M., Maccacaro T., Schild R., Wolter
A., Fleming, T.A., Henry, J.P., 1991, ApJS, 76, 813  

\bibitem[]{31}
Stoughton C., et al., 2002, AJ, 123, 485

\bibitem[]{32}
Wang T.G., Lu Y.J., Zhou Y.Y., 1998, ApJ, 493, 1

\bibitem[]{33}
York D.G., et al., 2000, AJ, 120, 1579

\end{thebibliography}
\end{document}